\documentstyle[prl,aps,twocolumn]{revtex}

\begin{document}

\title{THE UPPER CRITICAL DIMENSION OF THE KPZ EQUATION}

\author{Michael L\"assig$^{1}$ and Harald Kinzelbach$^{2}$}

\address{
$^{1}$ Max-Planck-Institut f\"ur Kolloid- und Grenzfl\"achenforschung,
Kantstr.~55, 14513 Teltow, Germany\\
$^{2}$ Universit\"at Heidelberg, Institut f\"ur theoretische Physik,
Philosophenweg~19, 69120 Heidelberg, Germany}

\date{August 7, 1996}
\maketitle

\begin{abstract}
The  strong-coupling regime of Kardar-Parisi-Zhang surface
growth driven by short-ranged noise has an upper critical dimension
$d_>$ less or equal to four (where the dynamic exponent $z$ takes the
value $z (d_>) = 2$). To derive this, we use the mapping onto
directed polymers with  quenched disorder. Two such polymers coupled by
a small contact attraction of strength $u$ are shown to form a bound
state at all temperatures $\beta^{-1} \! \leq \, \beta_c^{-1}$, the
roughening temperature of a single polymer.  Comparing the
singularities of the localization length $\xi_\perp \sim u^{1 / (2 -
d)}$ at $\beta_c^{-1}$ and  $\xi_\perp \sim u^{1 / 2 (1 - z)}$ below
$\beta_c^{-1}$ then yields $d_> \leq 4$.

\vspace{10 pt}
\vspace{24pt}
\end{abstract}


\vfill
\narrowtext 

The Kardar-Parisi-Zhang equation~\cite{KPZ} has been introduced as the
simplest nonlinear evolution equation
\begin{equation}
\partial_t h({\bf r},t)  =   \nu \, \nabla^2 h({\bf r}, t)
                         + \frac{\lambda}{2} \, ( \nabla h({\bf r},t))^2 \,
                         + \eta ({\bf r},t)
\label{KPZ}
\end{equation}
for a continuous ``height field'' $h( {\bf r}, t)$
driven by Gaussian noise $\eta ({\bf r},t)$ with
$\langle \eta({\bf r},t) \rangle = 0$ and
$\langle \eta({\bf r},t) \eta({\bf r}' ,t')\rangle =
\sigma^2 \delta^d ({\bf r}- {\bf r}') \delta(t-t')$.
It appears ubiquitously in a number of nonequilibrium statistical
problems such as fluid dynamics or dissipative transport,
as well as in systems with quenched disorder, for example flux lines in a dirty
type-II superconductor (see~\cite{reviews.KPZ} for reviews).

The morphology of a growing surface governed by~(\ref{KPZ}) is well
known in low spatial dimensions. The asymptotic scaling of the
spatio-temporal height correlations,
\begin{equation}
\langle ( h({\bf r}_1, t_1) - h({\bf r}_2, t_2) )^2 \rangle
    \sim r^{2 \chi} \,
    {\cal C} (t / r^z )
\end{equation}
(with $r \equiv |{\bf r}_1 - {\bf r}_2|$ and $ t \equiv |t_1 - t_2|$)
 defines the {\em roughness exponent} $\chi$ and the {\em dynamic
exponent} $z$. In the Gaussian dynamics ($\lambda = 0$), one has $\chi
= (2 - d)/2$ and $z = 2$. For $\lambda \neq 0$, Galilei invariance
imposes the relation $\chi + z = 2$ \cite{FisherHuse.paths}.  In
dimensions $d \leq 2$, any small nonlinearity $(\lambda/2) (\nabla
h)^2$ is a relevant perturbation of the Gaussian theory and leads to
new values of the exponents ($\chi = 1/2$, $z = 3/2$ in $d =
1$~\cite{FNS} and $\chi \approx 0.386$, $z \approx 1.612$ in $d =
2$~\cite{Ala-NissilaAl}). In the renormalization group, there is a
crossover between the Gaussian fixed point, which is
(infrared-)unstable, and the {\em strong-coupling} fixed point, which
is stable. For $d > 2$, a Gaussian surface is smooth. A small
nonlinearity $(\lambda/2) (\nabla h)^2$ does not alter this asymptotic
scaling. There is now a roughening transition to the strong coupling
regime at finite values $\pm
\lambda_c$~\cite{ImbrieSpencer,CookDerrida,EvansDerrida}.  In the
renormalization group, the transition is represented by a third fixed
point. This {\em critical} fixed point is unstable and appears between
the Gaussian fixed point and the strong-coupling fixed point which are
now both stable~\cite{TangNattermannForrest,FreyTauber}. The critical
point is characterized by the dimension-independent critical exponents
$\chi^\star = 0$ and $z^\star = 2$~\cite{DotyKosterlitz,Lassig.KPZ}.

It is a notorious difficulty of the strong-coupling regime that above
dimension one its properties are inaccessible to any known systematic
approximation, let alone to an exact solution. In particular,
renormalized perturbation theory with an $\varepsilon$-expansion about
$d = 2$ fails to produce a strong-coupling fixed
point~\cite{Lassig.KPZ}. This feature, which the Kardar-Parisi-Zhang
equation shares with other driven systems  such as fully developed
turbulence, is a major open problem in nonequilibrium statistical
physics. In this letter, we discuss a (modest) step towards its
eventual solution. We find that the Kardar-Parisi-Zhang equation has an
{\em upper critical dimension} $d_> \leq 4$, where the
exponents in the strong-coupling regime take the values  $z = 2$ and
$\chi = 0$, equal to those governing the limit $d \to
\infty$~\cite{d=infty,MezardParisi}
(and thus presumably the whole interval $d_> \leq d <
\infty$). As $d_>$ is approached from below, the exponents tend to
these values continuously. Hence, the upper critical dimension could
serve as the starting point for a controlled expansion. However, the
reader should be cautioned that the name ``upper critical dimension'' may
be misleading since $d_>$ does not mark the borderline to simple
mean-field behavior as in the standard theory of critical phenomena,
but instead to an even more complicated state in high dimensions with
presumed glassy characteristics~\cite{MezardParisi,MooreAl.modecoupling}.

Even the existence of a finite upper critical dimension has been very
controversial. Numerical work seems to indicate that a strong coupling
phase with nontrivial exponents $z<2$, $\chi >0$ persists in dimensions
$d=3,4, \dots, 7$ \cite{Ala-NissilaAl}.  Various theoretical arguments,
on the other hand, favor the existence of a finite upper critical
dimension $d_>$. This is the case for functional renormalization group
calculations~\cite{HalpinHealey.NattermannLeschhorn}, which are
supported by a Flory-type argument~\cite{FeigelmanAl}, and for a
$1/d$-expansion
based on a solution on the Cayley-tree \cite{CookDerrida} (criticized,
however, in \cite{FisherHuse.paths}). Recent work
treating the Kardar-Parisi-Zhang dynamics in mode-coupling
approximation gives contradictory results, predicting values of
$d_>$ between 3 and 4~\cite{modecoupling,MooreAl.modecoupling}
or $d_> = \infty$\cite{Tu}.

In contrast to all these approaches, the arguments of this Letter are
not based on any uncontrolled approximation scheme. We use the known
equivalence of the Kardar-Parisi-Zhang equation to a system of directed
polymers and compare {\em exact} (and individually testable) physical
properties of this system in the strong-coupling phase and at the
roughening transition. The importance of $d = 4$ for the roughening
critical point has already been stressed
recently~\cite{Lassig.KPZ,BundschuhLassig.KPZ}.  For $2 \leq d \leq 4$,
this universality class is accessible by renormalized perturbation
theory, which produces the exponents $\chi^\star = 0$ and $z^\star = 2$
exactly to all orders. At $d = 4$, however, there are singularities in
some observables (for example, in the exponent $y^\star$ defined
below).

Via the well-known Hopf-Cole transformation,
\begin{equation}
\exp \left [ \frac{\lambda}{2 \nu} h ({\bf r}_f ,t_f) \right ] =
\int {\cal D} {\bf r} \, \delta ( {\bf r}(t_f) - {\bf r}_f)
\exp \left [ - \beta  {\cal H} \right]
\label{Z}
\end{equation}
with the Hamiltonian
\begin{equation}
{\cal H} = \int_0^{t_f} {\rm d} t
           \left [ \frac{1}{2} \left(  \frac{{\rm d}{\bf r}}{{\rm d} t}
                                \right )^2 -
                    \lambda \eta({\bf r} (t), t) \right ] \;,
\label{H}
\end{equation}
the Kardar-Parisi-Zhang equation can be mapped onto the equilibrium problem
of a directed polymer ${\bf r}(t)$ living in the quenched random potential
$\lambda \eta({\bf r}, t)$ at temperature $\beta^{-1} = 2 \nu$.
The polymer is characterized by its transversal displacement
\begin{equation}
\overline{ \langle ( {\bf r}(t_1) - {\bf r}(t_2))^2 \rangle }
\sim |t_1 - t_2|^{2 \zeta}
\label{Delta}
\end{equation}
and by free energy quantities like the ``Casimir'' term
\begin{equation}
\overline f_c (R) \equiv
\lim_{T \to \infty} \partial_T( \overline F(T,R) - \overline F(T, \infty) )
\sim R^{(\omega - 1) / \zeta}
\label{fc}
\end{equation}
in a system of longitudinal size $T$ and transversal size $R$.
(Averages over the disorder are denoted by overbars,
thermal averages by brackets $\langle \dots \rangle$.)
The asymptotic scaling in (\ref{Delta}) and (\ref{fc}) is related to the
growth exponents by $\zeta = 1/z$ and $\omega = \chi/z$; the scaling
relation due to Galilei invariance now reads $\omega = 2 \zeta - 1$.
In the strong-coupling regime for $d < d_>$, the polymer
becomes superdiffusive ($\zeta > 1/2$), and its free energy acquires an
anomalous dimension $\omega > 0$. These disorder-induced fluctuations
persist in the zero-temperature limit, that is, in the ensemble of minimum
energy
paths $r_0(t)$. In the weak-coupling
(high-temperature) regime for $d > 2$, thermal fluctuations dominate
($\zeta = 1/2$) and hyperscaling is preserved ($\omega = 0$).
The roughening transition between these two phases takes place at a
finite temperature $\beta_c^{-1}$. At $d = d_>$ (and probably
for $d \geq d_>$), the
exponents $\zeta = 1/2$ and $\omega = 0$ govern the low-temperature
phase as well.

In its mapping to directed polymers, it is a natural extension of the
model to study an ensemble of lines $r_i(t)$ that live in the same random
potential and are coupled by direct mutual forces \cite{Mezard,Tang,fermions}.
The simplest case is that of just two lines with the interaction
\begin{equation}
{\cal H}_I = - u \int {\rm d} t \, \Psi (t)
\label{Hi}
\end{equation}
in terms of the local pair field
$\Psi (t) \equiv \delta ({\bf r}_1 (t) - {\bf r}_2(t))$.
For $d > 2$, this interaction is irrelevant at the Gaussian fixed point
governing the weak-coupling regime. However, for $d < d_>$, it turns
out to be relevant both at the strong-coupling fixed
point~\cite{fermions} and at the transition fixed point.  Thus for all
temperatures $\beta^{-1} \leq \beta_c^{-1}$, an
 attractive contact interaction ($u > 0$) leads to a bound state,
suppressing the thermally activated relative fluctuations of the lines
on scales
$|{\bf r}_1 - {\bf r}_2|
\,\raisebox{-.6ex}{$\stackrel{\displaystyle >}{\sim}\,$} \xi_\perp $.
The transversal localization length $\xi_\perp$ has the singularities
\begin{equation}
\xi_\perp \sim u^{- \zeta/y} \;\; \mbox{with} \;\;
y = 1 - \omega
\label{xi.sc}
\end{equation}
in the low-temperature phase ($\beta^{-1} < \beta_c^{-1}$), and
\begin{equation}
\xi_\perp \sim u^{- 1 / 2 y^*} \;\; \mbox{with} \;\;
y^* = (d - 2)/2 \;.
\label{xi.rt}
\end{equation}
at the critical temperature $\beta_c^{-1}$.  Of course, for any $u > 0$
kept fixed, $\xi_\perp$ should increase with temperature, since the
amount of entropic fluctuations increases.  This implies $ y / \zeta \geq
2 y^* $, that is, an upper bound on the free energy exponent, $ \omega
\leq (4 - d)/d $. Our main result $ d_> \leq 4$ then follows
immediately. We now discuss the arguments leading to
Eqns.~(\ref{xi.sc}) and (\ref{xi.rt}) above.

{\em (a) Strong coupling phase:}
At low temperatures and for $\omega > 0$, two noninteracting lines ($u = 0$)
in the same random potential have a stationary pair distribution
$P(r) \equiv \lim_{t, T - t \to \infty}
  \overline{ \langle \delta( {\bf r}_1(t) - {\bf r}_2(t) - {\bf r}) \rangle }$
given by
\begin{equation}
P(r) = \beta_R^{-1} r^{\theta} \;\;
\mbox{with} \;\; \theta = - d - \omega/\zeta
\label{Pcrit}
\end{equation}
in a suitable normalization~\cite{HwaFisher.paths,fermions}. The nonintegrable
short-distance singularity is cut off on scales
$r \,\raisebox{-.6ex}{$\stackrel{\displaystyle <}{\sim}\,$} \tilde \xi_\perp$,
where
$\tilde \xi_\perp \sim \beta_R^{- \zeta/\omega}$ is the crossover length to the
asymptotic strong-coupling behavior~\cite{betaR}. Thus with finite
probability, the two lines share a common ``tube'' of width $\tilde
\xi_\perp$ along the minimum energy path $r_0(t)$ for a single line;
this probability approaches one in the limit $\beta_R^{-1} \to 0$. In
that sense, the ground state $r_0(t)$ is {\em
unique}~\cite{uniqueness}.  At finite temperatures, however, the lines
make thermally activated individual excursions on all length scales
from the tube $r_0(t)$. These excursions generate the power law
distribution (\ref{Pcrit}), all positive integer moments of which
diverge.

Additional interactions between the lines probe the uniqueness of the
ground state as well as the statistics of the low energy
excursions~\cite{rsb}.  For example, a mutual repulsion (\ref{Hi}) with
$u < 0$ forces one of the lines onto a distant excited path. (In
superconductors, this is important to stabilize a dilute ensemble of
flux lines in the strong-coupling regime against collapse.) For the
purpose of this Letter, it is more useful to study a weak attractive
interaction ($u > 0$), which localizes the two lines to each other. The
normalized bound state distribution has the form $P(r, \xi_\perp) =
   N^{-1} (\xi_\perp) P(r) {\cal F}(r /
   \xi_\perp)$; on scales $r
\,\raisebox{-.6ex}{$\stackrel{\displaystyle >}{\sim}\,$} \xi_\perp$,
it falls off exponentially as given by the scaling function ${\cal F}$,
but in the scaling regime $r \ll \xi_\perp$, its dependence on
$\xi_\perp$ originates only from the overall normalization
$N(\xi_\perp) \equiv
   \int {\rm d}^d r P(r) {\cal F}(r / \xi_\perp)$.

Pair interactions have been treated  in ref.~\cite{fermions} by
renormalized perturbation theory for the Hamiltonian (\ref{Hi}), based
on the short-distance expansion
$\Psi(t) \Psi(t') \sim \beta_R^{-1} |t -t'|^{- \omega} \Psi (t)$
at the strong-coupling fixed point of noninteracting lines ($u = 0$).
For $u > 0$, one finds a bound state with the localization length
singularity (\ref{xi.sc}). This is in agreement with results in $d = 1$
from numerical work~\cite{Mezard,Tang} and from the dynamic
renormalization of an extended Kardar-Parisi-Zhang
equation~\cite{Mukherji}.

It is instructive to cast the field-theoretic derivation of
ref.~\cite{fermions} into the form of a variational scaling argument:
a weak bound state has a localization length that minimizes its free
energy per unit of $t$, balancing the free energy gain $\delta
\overline f_u$ from the overlap of the two lines with the loss $\delta
\overline f_c$ due to the confinement of their relative fluctuations
\cite{scalingargument}.  The overlap free energy is proportional to the
stationary expectation value of the pair field,
\begin{eqnarray}
\delta \overline f_u = - u \, \overline{ \langle \Psi \rangle }
 & \sim & - u N^{-1} (\xi_\perp)
\nonumber \\
& \sim & -u [1 + O (\beta_R^{-1} \xi_\perp^{- \omega / \zeta}) ] \;.
\label{deltafu}
\end{eqnarray}
The confinement energy has to vanish in the limit $\beta_R^{-1} \to 0$,
where all relative fluctuations of the lines are suppressed even
without attractive forces. Therefore, instead of the
temperature-independent Casimir energy $\xi_\perp^{(\omega - 1) / \zeta}$
analogous to (\ref{fc}), one expects
\begin{equation}
\delta \overline f_c \sim \beta_R^{-1} \xi_\perp^{-1/\zeta} \;,
\label{deltafc}
\end{equation}
a term that is analytic in the scaling variable $\beta^{-1}_R$ and
respects hyperscaling \cite{Casimir}.  The variation of $\delta
\overline f_u + \delta \overline f_c$ with respect to $\xi_\perp$ then
leads to (\ref{xi.sc}).

{\em (b) Roughening transition:}
In contrast to the strong-coupling fixed point, the renormalization
group for the critical fixed point in $2 < d < 4$ is well understood.
The polymer partition function (\ref{Z}) has been shown to be
{\em one-loop renormalizable}~\cite{Lassig.KPZ,BundschuhLassig.KPZ}.
It is this property that produces the exact dimension-independent exponents
$\chi^\star = 2$ and $z^\star = 2$; these values agree with previous results
from numerical work in $d = 3$~\cite{ForrestTang}, from dynamic renormalization
group calculations to one-loop~\cite{TangNattermannForrest} and
two-loop~\cite{FreyTauber} order, and from scaling
arguments~\cite{DotyKosterlitz}. At the critical point $\lambda_c$,
small variations of $\lambda$ are a relevant perturbation of dimension
$y^\star = (d - 2)/2$ (with $t$ as the basic scale)~\cite{Lassig.KPZ}. The
local
field conjugate to $\lambda$,
$\Phi_\eta (t) \equiv
  \int {\rm d}^d {\bf r}' \, \eta({\bf r}', t) \delta^d ({\bf r} (t) - {\bf
r}')$,
encodes the random potential evaluated along the polymer paths. This
field generates the crossover from the critical fixed point to the
strong-coupling fixed point ($\lambda \to \infty$) and to the Gaussian
fixed point ($\lambda = 0$). It is now a purely formal matter to
relate its disorder-averaged correlation functions to those of the
pair field $\Psi$ for $u = 0$~\cite{fermions,depinning}.  The simplest
example is the stationary one-point function
$\overline{ \langle \Phi_\eta \rangle } (R)$
obtained from the free energy per unit of $t$,
$ \overline f (R) = - \beta^{-1}
  \lim_{T \to \infty} \partial_T
  \overline{ \log {\rm Tr} \exp (- \beta {\cal H}) } (T,R)$.
With the disorder correlation and the pair
interaction regularized on the microscopic scale $a$,
$\overline{ \eta({\bf r}', t') \eta ({\bf r}'', t'') } =
  \sigma^2 \delta_a ({\bf r}' - {\bf r}'') \delta (t' - t'')$ and
$\Psi_a (t) \equiv \delta_a ({\bf r}_1 (t) - {\bf r}_2 (t))$,
the universal parts~\cite{contactterm} of
$\overline{\langle \Phi_\eta (t) \rangle} = \overline{\langle \Phi_\eta
\rangle}$
and $\overline{\langle \Psi_a \rangle}$ are seen to be proportional:
\begin{eqnarray}
\lefteqn{ \lambda^{-1} \overline{ \langle \Phi_\eta \rangle }
          = - \lambda^{-1} \partial_\lambda \overline f }
\nonumber
\\ & &
  = - \lambda^{-2} \sigma^2 \int\! {\rm d}^d {\bf r}' {\rm d}^d {\bf r}''
  \delta_a ({\bf r}' - {\bf r}'')
  \overline{ \frac{\delta^2 f}{\delta \eta ({\bf r}', t)
                               \delta \eta ({\bf r}'', t)} }
\nonumber
\\ & &
  =  \beta \sigma^2 \int\! {\rm d}^d {\bf r}' {\rm d}^d {\bf r}''
  \delta_a ({\bf r}' - {\bf r}'')
  \overline{ \langle \delta ({\bf r} (t) - {\bf r}')
                     \delta ({\bf r} (t) - {\bf r}'') \rangle^c }
\nonumber
\\ & &
  = \! - \beta \sigma^2 \! \int\! {\rm d}^d {\bf r}' {\rm d}^d {\bf r}''
  \delta_a ({\bf r}' - {\bf r}'')
  \overline{ \langle \delta ({\bf r} (t) - {\bf r}')  \rangle
             \langle \delta ({\bf r} (t) - {\bf r}'') \rangle }
\nonumber
\\ & &
  = - \beta \sigma^2 \, \overline{ \langle \Psi_a \rangle } \;.
\label{Psia}
\end{eqnarray}
At $\lambda = \lambda_c$, the correlation functions of $\Phi_\eta$ are
scale-invariant with the exponent $ x^\star = 1 - y^\star = (4 - d)/2$.
By (\ref{Psia}), the same holds for the correlation functions of
$\Psi_a$; for example,
\begin{equation}
\overline{ \langle \Psi_a \rangle } (R, \lambda_c) \sim
\overline{ \langle \Phi_\eta \rangle } (R, \lambda_c) \sim
    R^{ - x^\star / \zeta^\star } \;.
\label{onepoint}
\end{equation}
Hence, the local pair interaction $\Psi_a(t)$ is like $\Phi_\eta (t)$ a
relevant
scaling field of dimension $x^\star$ at the critical fixed point.
This result is consistent with the one-loop dynamic renormalization
group discussed in ref.~\cite{Mukherji}. We conclude that
the bound state generated by this interaction has the
localization length singularity (\ref{xi.rt}) at
$\beta_c^{-1}$~\cite{thermalsing}.  Just below that temperature, there
is a crossover from (\ref{xi.rt}) to (\ref{xi.sc}) at the scale $u \sim
\beta_c^{-1} - \beta^{-1}$.

The singularity (\ref{xi.rt}) can again be obtained from a variational
scaling argument. At $\beta_c^{-1}$, the competing free energy contributions
analogous to (\ref{deltafu}) and (\ref{deltafc}) are the overlap term
\begin{equation}
\delta \overline f_u
  = - u \overline{ \langle \Psi_a \rangle }
  \sim - u \xi_\perp^{ - x^\star / \zeta^\star }
  = - u \xi_\perp^{d - 4}
\end{equation}
following from (\ref{onepoint}) and a Casimir term of the form (\ref{fc}),
\begin{equation}
\delta \overline f_c \sim \xi_\perp^{ (\omega^\star - 1) / \zeta^\star }
                     = \xi_\perp^{-2} \;.
\end{equation}

It is tempting to speculate about the nature of the strong-coupling
regime in high dimensions. Below $d_>$, the uniqueness of the ground state
implies a collapse of the pair distibution function $P(r)$ in the limit
$\beta_R \to 0$, see Eq.~(\ref{Pcrit}). This property may be lost in a
glassy phase. If many states remain accessible to the system,
long-ranged pair correlations $P(r) \sim r^\theta$ may persist even as
$\beta_R \to 0$. Thus certain characteristics of the strong-coupling
regime may qualitatively resemble those at the roughening transition
below $d_>$.


\begin{references}
\bibitem{betaR}
The ``renormalized'' temperature $\beta_R$~\cite{depinning}, a nonlinear
function of the bare continuum variables $\beta^{-1}, \lambda, \sigma^2$
in (\ref{Z}) and (\ref{H}), is an irrelevant scaling variable of
dimension $ - \omega$ at the strong-coupling fixed point.  If the
theory is defined on a lattice with lattice constant $a$, it is to be
identified with the physical temperature in the strong-coupling limit
$\tilde \xi_{||} \sim a$.
\bibitem{Casimir}
This modification can be understood heuristically by noting that the
Casimir pressure originates essentially from critical fluctuations
being cut off on the scale $r \sim \xi_\perp$. Compared to the case of a
single line~\cite{scalingargument}, it should therefore be reduced by a
factor
$P(r = \xi_\perp, \xi_\perp) / P_1 (r = \xi_\perp, \xi_\perp)
  \sim \xi_\perp^{-\omega / \zeta}$,
reflecting the reduced probability of a critical fluctuation of
size $r \sim \xi_\perp$.
\bibitem{contactterm}
In replacing the connected thermal correlator in the third line of
(\ref{Psia}) by the disconnected product in the forth line, we have
suppressed the full correlator
$\overline{ \langle \delta ({\bf r}(t) - {\bf r}')
                   \delta ({\bf r}(t) - {\bf r}'') \rangle }
 = \delta ({\bf r}' - {\bf r}'')$,
which contributes only the divergent contact term
$\beta \sigma^2 \delta_a (0)$ to the r.h.s..
\end{references}
\end{document}